\documentclass[aps,prd,twocolumn,superscriptaddress,nofootinbib]{revtex4}
\usepackage[utf8x]{inputenc}
\usepackage[T1]{fontenc}
\usepackage{lmodern}
\usepackage[british]{babel}
\usepackage{graphicx}
\usepackage{grffile}
\usepackage{hyperref}
\usepackage{color}
\usepackage{slashed}
\usepackage{textcomp}
\usepackage{subfigure}
\usepackage{amsmath}
\usepackage{amssymb}
\usepackage{amsfonts}
\usepackage{lscape}
\usepackage{psfrag}
\newcommand{\I}{\ensuremath{\mathrm{i}}}
\newcommand{\aetap}{\text{a--}\eta'}

\newcommand{\afn}{\text{a--}f_0}

\newcommand{\mps}{\ensuremath{m_{\text{PS}}}}
\newcommand{\mv}{\ensuremath{m_{\text{V}}}}
\newcommand{\ms}{\ensuremath{m_{\text{S}}}}
\newcommand{\mpcac}{\ensuremath{m_{\text{\tiny PCAC}}}}
\newcommand{\mpv}{\ensuremath{m_{\text{PV}}}}
\newcommand{\msh}{\ensuremath{m_{1/2}}}
\newcommand{\mgb}{\ensuremath{m_{0^{++}}}}
\newcommand{\mgbmp}{\ensuremath{m_{0^{-+}}}}
\newcommand{\su}[1]{\ensuremath{\text{SU}(#1)}}

\newcommand{\so}[1]{\ensuremath{\text{SO}(#1)}}
\newcommand{\uu}[1]{\ensuremath{\text{U}(#1)}}
\newcommand{\bpsi}{\bar{\psi}}

\newcommand{\shalf}{\ensuremath{\text{spin\,1/2\ }}}

\newcommand{\Pro}{\mathbb{P}}
\DeclareMathOperator{\Tr}{Tr}
\begin{document}

\title{The spectrum and mass anomalous dimension of SU(2) adjoint QCD with two
Dirac flavours}


\author{Georg Bergner}
\email{bergner@itp.unibe.ch}
\affiliation{Albert Einstein Center for Fundamental Physics,Institute for Theoretical Physics, University of Bern, Sidlerstr.~5, CH-3012 Bern, Switzerland}

\author{Pietro Giudice}
\email{p.giudice@uni-muenster.de}
\author{Gernot M\"unster}
\email{munsteg@uni-muenster.de}
\affiliation{University of M\"unster, Institute for Theoretical Physics, Wilhelm-Klemm-Str.~9, D-48149 M\"unster, Germany}
\author{Istvan Montvay}
\email{montvay@mail.desy.de}
\affiliation{Deutsches Elektronen-Synchrotron DESY, Notkestr. 85, D-22603 Hamburg, Germany}

\author{Stefano Piemonte}
\email{stefano.piemonte@ur.de}
\affiliation{University of Regensburg, Institute for Theoretical Physics, Universit\"atsstr.~31, D-93040 Regensburg, Germany}


\date{2nd June 2017}

\begin{abstract}
In this work we present the results of our investigation of \su{2} gauge
theory with two Dirac fermions in the adjoint representation (aQCD2), which
belongs to the class of strongly interacting gauge theories that are of
basic interest for extensions of the Standard Model. We have done numerical
lattice simulations of this theory at two different values of the gauge
coupling and several fermion masses. Our results include the particle
spectrum and the mass anomalous dimension. The spectrum contains new exotic
fermion-gluon states and flavour-singlet mesons. The mass anomalous
dimension is determined from the scaling of the masses and the mode number.
The remnant dependence of the universal mass ratios and mass anomalous
dimension on the gauge coupling indicates the relevance of scaling
corrections, such that earlier estimations for the universal fixed point
value of the mass anomalous dimension are incomplete without their
inclusion.
\end{abstract}
\maketitle

\section{Introduction}

New strongly interacting gauge theories are interesting possibilities for an
extension of the Standard Model of particle physics. This leads to the
general theoretical question about possible realisations of these
interactions and, in particular, whether a dynamics and a particle spectrum
completely different from QCD can be observed. These questions motivate the
investigation of SU(N) gauge theories with 
different numbers of fermions in different representations
of the gauge group. Particularly interesting are theories with
an infrared fixed point in the ``conformal window''.

The possible realisations of strong interactions that might be relevant for
extensions of the Standard Model are a motivation for our current
investigation of theories with fermions in the adjoint representation of the
gauge group. The adjoint representation is particularly interesting among
the higher representations of the gauge group. This representation is
employed in several interesting theories, including supersymmetric
Yang-Mills theory and Technicolor candidates. The objectives of our
investigations are, however, not phenomenological aspects of theories
extending the Standard Model, but to study basic non-perturbative
characteristics of the dynamics and structure of gauge theories different
from QCD.

One example for extensions of the Standard Model are Technicolor theories
\cite{Weinberg:1975gm,Susskind:1978ms}. They provide a more natural
representation of the electroweak sector by introducing a new strongly
interacting sector on a higher energy scale. The Higgs particle emerges as a
light scalar bound state of elementary particles in the new sector. The absence of
other bound states in the same mass region requires a mechanism for the
generation of a strong mass hierarchy with a light scalar. This can not be
achieved by a simple modification of standard QCD.

This non-QCD like feature, and other related ones, could be obtained as a
consequence of a ``walking'', i.\,e.\ near IR conformal behaviour of the gauge
coupling \cite{Holdom:1981rm}. The running of the gauge coupling typically
becomes slower with an increasing number of fermions, which is already evident
from the perturbative beta function. With a suitable set of fermions a
conformal window appears, in which the running terminates at an infrared
fixed point, where the theory becomes scale independent. The upper boundary
of the conformal window, where the fixed point disappears due to the loss of
asymptotic freedom, is determined by the perturbative running in the weak
coupling regime, whereas the lower boundary can only be investigated by
non-perturbative methods. Several analytical
\cite{Sannino:2004qp,Dietrich:2006cm,Braun:2010qs} and numerical lattice
studies
\cite{Catterall:2007yx,Hietanen:2009zz,DelDebbio:2010hx,DeGrand:2011qd,Appelquist:2011dp,DeGrand:2010na,Fodor:2016wal,Brower:2015owo,Athenodorou:2014eua}
have been dedicated to the investigation of the conformality of different
gauge theories. For a review concerning the lattice results see
\cite{Appelquist:2016viq,DeGrand:2015zxa,Nogradi:2016qek,Giedt:2015alr}.

Theories with fermions in higher representations are favoured in this
approach since they allow for a near IR conformal behaviour with a relatively
small number of fermions. In particular, the theory with $N_f=2$ Dirac
fermions in the adjoint representation of \su{2}, also called Minimal
Walking Technicolor, has interesting applications in phenomenological
models \cite{Foadi:2007ue}. Apart from that, the questions about the
size of the conformal window and theories with different realisations of
strong interactions are of basic quantum field theoretic interest.

Other interesting gauge theories with fermions in the symmetric and
anti-symmetric tensor representation are related to models in the adjoint
representation by large $N_c$ equivalence. This leads to constraints for the
conformal window of models in the symmetric representation that can be
deduced from the adjoint one \cite{Bergner:2015dya}.

In this work we present our results for \su{2} gauge theory with $N_f=2$
Dirac fermions in the adjoint representation (adjoint QCD, aQCD2), including
a comparison to our previous studies of supersymmetric Yang-Mills theory,
which corresponds effectively to a Dirac
fermion flavour number $N_f = 1/2$. 
We
focus on the near IR conformal behaviour, the appearance of a light scalar
particle, and a large mass anomalous dimension. Compared to other studies of
such models, our results include the investigation of particle states that
have not been considered so far, namely new exotic fermion-gluon bound
states that are special in this theory, and mesonic states in the flavour
singlet channel. Moreover, we have determined the mass anomalous dimension
with a new method similar to the one introduced in \cite{Fodor:2016hke} for
the determination of the mode number. Our results on the dependence of the
anomalous dimension on the bare gauge coupling give new evidence for the
relevance of scaling corrections.

The existence of an infrared fixed point is a universal feature of a given
theory, but the direct determination of the conformality from the running of
the coupling might be biased by technical difficulties and the scheme
dependence. An alternative approach for the determination of universal
properties like the existence of the fixed point and the mass anomalous
dimension is the investigation of mass deformed theories. The conformal
behaviour of such a theory manifests itself in the particle spectrum. In the
first approximation the masses $M$ of all states should scale to zero
according to $M\propto m^{1/(1+\gamma^\ast)}$, where $m$ is the residual
quark mass, and the mass anomalous dimension $\gamma^\ast$ is the same for
all states \cite{DelDebbio:2010jy}. This hyperscaling should be observable
if $m$ is below a certain threshold. It is quite different from the chiral
symmetry breaking scenario, where a clear separation between the
pseudo-Nambu-Goldstone bosons (pNGb) and the rest of the spectrum appears at
small $m$, and eventually the mass of the pNGb goes to zero in the chiral
limit, whereas the masses of the other particles remain finite. It is in
general difficult to discern to which of the two classes the considered
theory belongs, since one is always restricted to a limited range of $m$ in
the lattice simulations, and the chiral limit $m=0$ can only be reached by
extrapolation. An additional difficulty is the influence of the gauge
coupling, which is expected to be irrelevant at the infrared fixed point,
but can in principle still be nearly marginal, i.~e.\ the scaling exponent
is $y_0 \lesssim 0$. The inclusion of the related scaling corrections has
been the subject of recent investigations \cite{Cheng:2013xha}. It turned
out that the inclusion of these correction was essential to arrive at
universal results from simulations with different lattice actions. In our
current work we make first investigations of the significance of these
scaling corrections. The investigation of these effects is important since
they might explain the differences in various estimations of the universal
behaviour of Technicolor candidates. A complete analysis of these effects
would, however, require a larger number of simulations.

In a QCD-like theory asymptotic freedom implies that the continuum limit is
reached at vanishing gauge coupling. The dependence of physical quantities
on the gauge coupling is hence essential to determine the universal
continuum limit. On the other hand, the gauge coupling is irrelevant in a
theory in the conformal window, as long as it is not in the strong coupling
regime. However, the position and the existence of the infrared fixed point
are not known a priori and hence one cannot rely beforehand on these
assumptions. Even if the $\beta$-function has a non trivial zero,
corrections to the scaling behaviour appear in the weak coupling regime,
where the effects of the lattice cut-off disappear. Our results show that
indeed the scaling properties of adjoint QCD on the lattice do depend on
gauge coupling. In particular, the mass anomalous dimension $\gamma^\ast$ is
lower if extracted from our ensembles with larger $\beta$.

A comparison with supersymmetric Yang-Mills theory, which is clearly below
the conformal window, might help to resolve the differences between the
conformal and the chiral symmetry breaking scenario. In such a comparison it
is important to choose a comparable lattice realisation since lattice
artefacts might have a significant influence on the scaling behaviour.

This paper is organised as follows: in Section~\ref{sec:gadjQCD} we present
an overview of the general chiral symmetry pattern and the continuum
formulation of adjoint QCD. In Section~\ref{sec:LMWT} we present our general
setup for the numerical investigations of aQCD2 including the considered
lattice action. In Section~\ref{sec:parameters} we discuss the range of the
simulation parameters. Section~\ref{sec:specMWT} summarises our numerical
results for the particle spectrum of this theory with a focus on the most
important states, the glueball, the fermionic \shalf state, and the scalar
singlet meson. This includes details about the uncertainties in the
numerical estimations. Estimates for the mass anomalous dimension from the
particle spectrum and the mode number are provided in Section~\ref{sec:gam}.
We also include a short explanation of the method for the mode number
estimation since it is different from the one used in earlier investigation
of this theory. In Section~\ref{sec:concl} we finally discuss implications
of our results and possible directions for further investigations.

\section{Chiral symmetry breaking scenario and conformality in adjoint QCD}
\label{sec:gadjQCD}

The theory investigated in this work is \su{2} adjoint QCD with $N_f=2$
Dirac fermions (aQCD2). The Lagrangian of adjoint QCD has the following form
\begin{equation}
\mathcal{L}=
\Tr\left[-\frac{1}{4}F_{\mu\nu}F^{\mu\nu}
+\sum_{i=1}^{N_f}\bpsi_{i}(\slashed{D}+m_0)\psi_{i}\right].
\end{equation}
Here $\psi$ is a  Dirac-Fermion in the adjoint representation of \su{2} with
the covariant derivative
\begin{equation}
D_\mu \psi = \partial_{\mu}\psi + \I g [A_{\mu},\psi]\,.
\end{equation}

The adjoint representation is consistent with the Majorana condition
$\lambda=C\lambda^T$, which means that each Dirac fermion $\psi_k$ can be
decomposed into two Majorana fermions $\lambda_i$, and the two Majorana
flavours are not mixed by the action. In particular using
$\psi_k=\frac{1}{\sqrt{2}}(\lambda_{2k}+\I\lambda_{2k+1})$ one gets
\begin{equation}
\mathcal{L}=
\Tr\left[-\frac{1}{4}F_{\mu\nu}F^{\mu\nu}
+\frac{1}{2}\sum_{k=1}^{2N_f}\bar{\lambda}_{k}(\slashed{D}+m)\lambda_{k}\right].
\end{equation}
In this way, theories with an odd number of Majorana flavours can be
considered to have half integer Dirac flavours.

Chiral symmetry breaking results from the formation of a condensate in the
massless theory or from a mass term. In our present case the breaking
pattern is different from QCD and related to the transformation properties
of the Majorana flavours. The left handed and right handed Weyl components
of the $2N_f$ Majorana field are, however, not independent since they are
related by the Majorana condition. Considering the action formulated in
terms of Majorana fermions in the Weyl representation one observes the
chiral symmetry breaking pattern \cite{Munster:2014cja}
\begin{align}
\su{2N_f} \rightarrow \so{2N_f}\,.
\label{chiralbreak}
\end{align}
As a consequence, there are $2N_f^2+N_f-1$ pseudo Nambu-Goldstone bosons
(pNGb) generated in adjoint QCD, if the chiral symmetry is broken by the
chiral condensate as in QCD. The chiral symmetry for the Dirac fermions,
$\su{N_f}\times\su{N_f}\times\uu{1}_V$ broken to $\su{N_f}_V\times\uu{1}_V$,
is of course included as a subgroup of the above $\su{2N_f}$. In particular,
the unbroken $\so{2N_f}$ contains always the vector-like $\uu{1}_V$ and the
same pseudoscalar mesonic states (pions), which can be formulated with Dirac
fermions for $N_f>1$, provide a signal for pNGb. For $N_f<2$ the operator
$\psi^T C \gamma_5 \psi$ describes a pNGb \cite{Athenodorou:2014eua}. This
signal can also be used for chiral symmetry breaking in supersymmetric
Yang-Mills theory considered in a partially quenched setup
\cite{Munster:2014cja}. In the chiral limit the spectrum is expected to be
separated into the light pNGb and the other heavier states.

If the theory is inside the conformal window, a completely different
behaviour of the spectrum is expected. In the conformal limit, where the
fermions become massless, there is no remnant mass scale. The beta function
approaches the infrared fixed point in this limit. Consequently all masses
scale to zero according to $M\propto m^{1/(1+\gamma^\ast)}$ with the
renormalised quark mass $m$ and the mass anomalous dimension $\gamma^\ast$
at the fixed point. The ordering of the different states is not determined
in this scaling relation. Nevertheless, one might expect a light scalar as
an approximate \textit{dilaton} due to the restoration of dilatation
symmetry in the conformal limit. Indeed, several investigations have found
indications for a light scalar in (near) IR conformal theories. Even though all
masses scale to zero, their ratios can be extrapolated to the conformal
limit. It seems that these ratios are a universal characteristic for each
(near) IR conformal theory \cite{Athenodorou:2016ndx}.

\section{Adjoint QCD on the lattice}
\label{sec:LMWT}

Our lattice formulation of the theory employs the Wilson gauge action built
from the plaquette variables $U_p$ and the Wilson-Dirac operator in the
adjoint representation. In its basic form the lattice action reads
\begin{multline}
\mathcal{S}_L =
\beta \sum_p\left(1-\frac{1}{N_c}\mbox{Re}\,\mathrm{tr} U_p\right)\\
+\sum_{xy,N_f} \bar{\psi}_x^{n_f}(D_w)_{xy}\psi_y^{n_f}\, ,
\end{multline}
where $D_w$ is the Wilson-Dirac operator
\begin{multline}
(D_w)_{x,a,\alpha;y,b,\beta}
    =\delta_{xy}\delta_{a,b}\delta_{\alpha,\beta}\\
    -\kappa\sum_{\mu=1}^{4}
      \left[(1-\gamma_\mu)_{\alpha,\beta}(V_\mu(x))_{ab}
                          \delta_{x+\mu,y} \right . \\
    \left. +(1+\gamma_\mu)_{\alpha,\beta}(V^\dag_\mu(x-\mu))_{ab}
                          \delta_{x-\mu,y}\right].
\end{multline}
The hopping parameter $\kappa$ is related to the bare fermion mass via
$\kappa=1/(2m_0+8)$, and the index $N_f$ runs over the number of different
fermion flavours, where here $N_f=2$.

The link variables $U_{\mu}(x)$ are in the fundamental representation of the
gauge group \su{2}. The adjoint gauge field variables $V_{\mu}(x)$ in the
Wilson-Dirac operator are the corresponding elements in the adjoint
representation. They are defined by $[V_\mu(x)]^{ab}=2\,\mathrm{tr} [
U^\dag_\mu(x) T^a U_\mu(x) T^b]$, where $T^a$ are the generators of the
gauge group normalised such that $2\,\mathrm{tr} [T^a T^b]=\delta^{ab}$.

The basic lattice action has been applied in earlier studies
\cite{Catterall:2007yx,DelDebbio:2010hx,DelDebbio:2015byq}. In our
simulations we use an improved version of this lattice action with a
tree-level Symanzik improved gauge action and stout smearing for the link
fields in the Wilson-Dirac operator \cite{Morningstar:2003gk}. It is
expected that these modifications reduce the lattice artefacts. In most of
our runs the stout smearing is iterated three times with the smearing
parameter $\rho=0.12$.

Our numerical lattice simulations have been performed with the two-step
polynomial hybrid Monte Carlo (PHMC) algorithm \cite{Montvay:2005tj}. This
algorithm is based on polynomial approximations of the inverse powers of the
lattice action. The first polynomial gives a crude approximation which is
corrected by the second polynomial. This correction is especially important
near zero fermion mass, where the inverse power has a singularity. In our
simulations we have chosen the second polynomial in such a way that the
lower bound of the approximation interval was by about a factor 10 smaller
than the smallest occurring eigenvalues. In this case the approximation is
already so good that in practice no further correction by a reweighting
factor is necessary.

\section{Simulation parameters and continuum limit}
\label{sec:parameters}

\begin{figure}[ht]
\centering\includegraphics[width=\columnwidth]{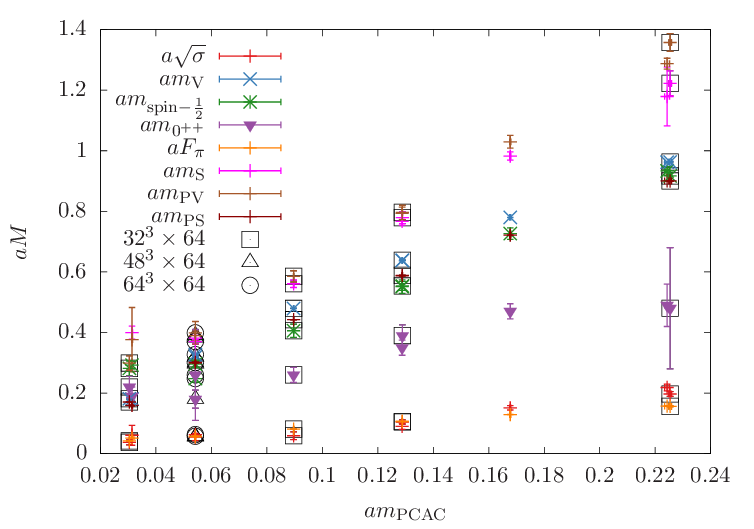}
\caption{The results for the mass spectrum of aQCD2 at $\beta=1.5$. The
masses, pseudoscalar decay constant, and string tension $\sigma$ as a
function of the renormalised fermion mass $\mpcac$ in lattice units, see
Table~\ref{tab:data} in Section~\ref{ap:data}.}
\label{fig:massres15a}
\end{figure}
\begin{figure}[ht]
\centering\includegraphics[width=\columnwidth]{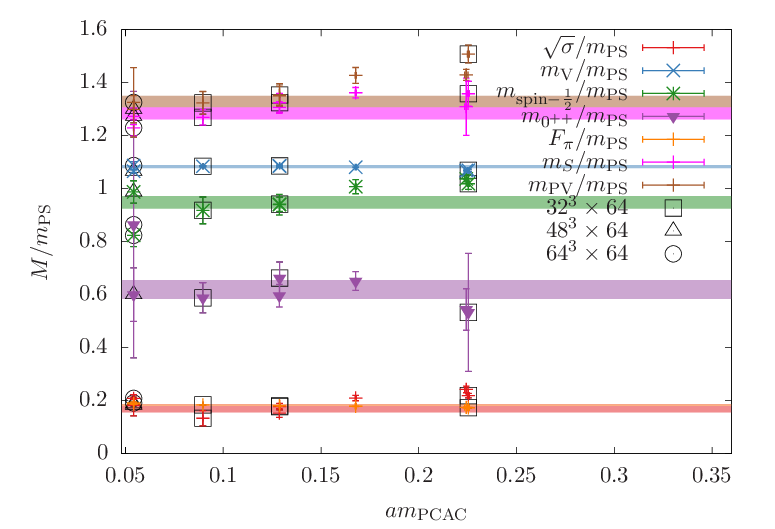}
\caption{The ratio between the different quantities and the pseudoscalar
meson mass at $\beta=1.5$. The plot includes a fit assuming an approximately
constant value of these ratios as a function of the fermion mass, see
Table~\ref{tab:massrat}.}
\label{fig:massres15b}
\end{figure}

In a confining gauge theory with a mass gap the lattice spacing is
determined by the coupling constant $\beta$. It can be defined in terms of a
scale setting quantity such as the Sommer scale $r_0$ or the string tension
$\sqrt{\sigma}$. Due to asymptotic freedom, the lattice spacing decreases as
$\beta$ is increased.

Close to the fixed point in an IR conformal gauge theory $\beta$ is an
irrelevant parameter which implies only a weak dependence on this parameter.
In the IR conformal theory the continuum limit can be defined only in terms of
the lattice spacing in units of the volume, the only remaining scale.
Nevertheless, the theories in the conformal window are still asymptotically
free and the continuum limit corresponds to the limit where $\beta$ goes to
infinity. Consequently a relevant dependence on $\beta$ is expected further
away from the infrared fixed point and closer to the Gaussian fixed point.
In a first approximation the $\beta$ dependence appears as a correction of
the scaling close to the fixed point. This has been investigated in a finite
size scaling analysis \cite{Cheng:2013xha}, where an agreement between
results from different lattice action has been possible in this way. The
connection between scaling corrections and discretisation errors has been
pointed out in \cite{Giedt:2015alr}.

A finite mass breaks conformal symmetry and implies further corrections to
the simple scaling picture. The mass term in the action is a relevant
parameter and the renormalisation group flow hence does not approach the
infrared fixed point. However the running of $\beta$ is still expected to be
rather weak at least for smaller masses. At a fixed mass, the influence of
lattice artefacts can be investigated by comparing different values of
$\beta$, where the largest value corresponds to the smallest lattice
spacing.

In fact, in numerical simulations it is for several reasons impossible to
reach the limit of an exactly vanishing fermion mass. In a conformal theory
this limit would introduce severe finite size effects and hence the
interpretation of the particle spectrum would be difficult. In addition, the
range of fermion masses is limited by the updating algorithm, where the cost
of the simulation rises exponentially if the fermion masses are approaching
zero.

The possible range of $\beta$ values in aQCD2 is constrained from below by
the bulk phase transition. The bare parameters of this transition in our
investigations are different from those in previous investigations, which is
related to the change of the gauge action. With our lattice action we
determined the position of the bulk transition to be around $\beta=1.4$. The
control of finite volume effects is important in the investigations of an IR
conformal theory. Therefore in our first analysis we have chosen
$\beta=1.5$, which is not much above the bulk transition. This allows
changing the lattice volume in a wide range. In a second step we have also
done simulations at $\beta=1.7$ to check for possible lattice artefacts and
scaling corrections.

The pseudoscalar meson mass in lattice units in these runs was in the range
between $0.9$ and $0.2$. Finite size effects are generally quite significant
in simulations of (near) IR conformal theories. We have found that at small
lattice volumes the ordering of the states is significantly changed. The
most relevant scale for the finite size effects is the mass of the lightest
bound state, in aQCD2 the $0^{++}$ glueball. The finite size effects lead,
however, to a larger mass of this particle, which makes the estimate
ambiguous. Therefore we have considered the mass of the pseudoscalar meson
which can be easily determined quite precisely.

\begin{figure}
\includegraphics[width=\columnwidth]{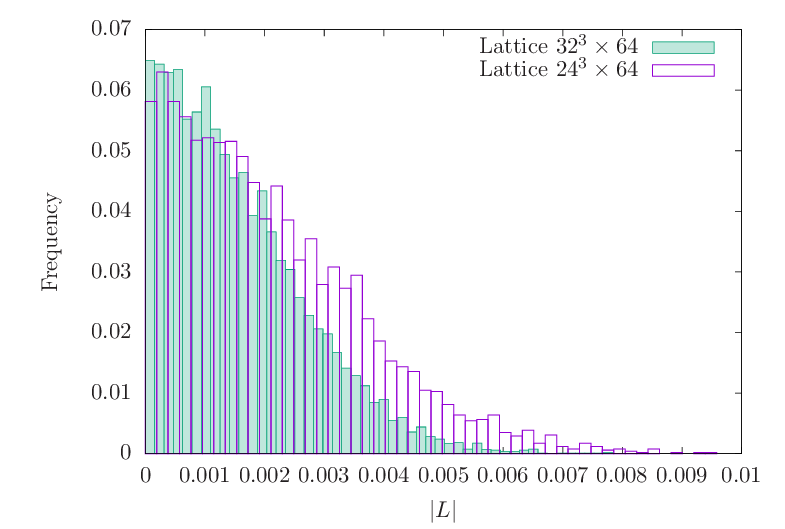}
\caption{This Figure shows a histogram of the absolute value of the Polyakov line in spatial direction for two different 
volumes ($\beta=1.5$, $\kappa=0.1350$). A peak at a nonzero value of this quantity 
would indicate the deconfinement phase transition. Like for all other considered parameters, the theory is 
in the confined phase. The increasing width at the smaller volume indicates a tendency towards the deconfinement transition.}
\label{fig:polyakovline}
\end{figure}

In general it is possible that there are also phase transitions and
indications of deconfinement at small box sizes. We have checked the
distribution of the Polyakov line at the smaller masses for any signal of a
transition. In none of the simulations we have found an indication for a
transition (see for example Figure~\ref{fig:polyakovline}), but at $\beta=1.5$, 
$\kappa=0.1350$ on the $24^3\times 64$
lattice there is a significant broadening of the spatial Polyakov line
distribution. Therefore we have excluded this run from the analysis.

A known difficulty of numerical simulations near the continuum limit is the
freezing of the topological charge of the gauge field in the finite physical
volume of the simulation (\textit{topology freezing}). This effect can be
made mild by choosing longer Hybrid Monte Carlo (HMC) trajectories
\cite{Meyer:2006ty,McGlynn:2014bxa}. Therefore, in most cases we ran the
PHMC updates by HMC trajectory of length 2, which mostly resulted in
acceptable integrated autocorrelation times of the topological charge. For
the smallest fermion masses, however, this autocorrelation time is
drastically increased to about 500 in HMC time, a value which is just by a
factor of 10 shorter than the total HMC time of the simulations. This
problem is present in the runs at ($\beta=1.5,\;\kappa=0.135$) and
($\beta=1.7,\;\kappa=0.13$). In these cases the sampling of different
topological sectors is poor. As shown below, these are the runs that are
also affected by large finite size effects and therefore they are not
considered for the main results.

\section{The lightest particles}
\label{sec:specMWT}

The primary focus of our investigation is the spectrum of lowest lying bound
states in adjoint QCD. The bound state spectrum consists of
mesonic states, glueballs, and mixed fermion-gluon states. We consider these
states as functions of the renormalised fermion mass, which we take to be
the PCAC mass $\mpcac$, determined through the partially conserved
axial-vector current relation. In addition to the particle masses, we have
determined the string tension $\sigma$ from the static quark-antiquark
potential and the pseudoscalar decay constant $F_\pi$.

The considered mesonic states include the pseudoscalar ones created by
$\bar{\psi_i}\gamma_5 \delta_{ij} \psi_j$ in the singlet and
$\bar{\psi_i}\gamma_5 \tau^{a}_{ij} \psi_j$, $a=1,2,3$, in the triplet
channel, where $\tau^{a}$ are the Pauli matrices, corresponding to the
adjoint eta prime meson ($m_{\aetap}$) and the pseudoscalar meson ($\mps$),
respectively. The triplet and singlet channel for the operator $\bar{\psi_i}
\psi_j$ correspond to the scalar meson ($\ms$) and the adjoint $f_0$
($m_{\afn}$) meson. In addition, also the vector meson ($\mv$), created by
$\bar{\psi_i}\gamma_k \psi_j$ ($k$ in spatial direction), and the
pseudovector meson ($\mpv$) in the triplet channel, created by
$\bar{\psi_i}\gamma_5 \gamma_k \psi_j$, have been considered. In the current
analysis we have also measured the scalar glueball ($\mgb$) and a mixed
fermion-gluon state with \shalf ($\msh$).

\begin{figure}[ht]
\centering\includegraphics[width=\columnwidth]{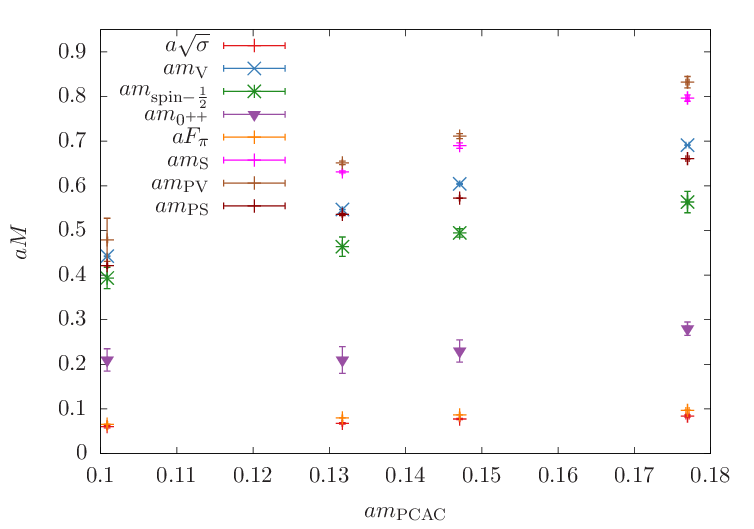}
\caption{As Figure~\ref{fig:massres15a} but with $\beta=1.7$.}
\label{fig:massres17a}
\end{figure}
\begin{figure}[ht]
\centering\includegraphics[width=\columnwidth]{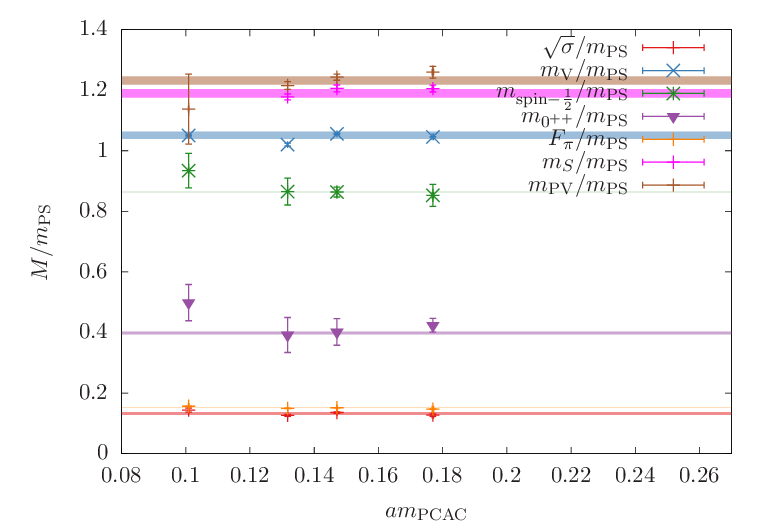}
\caption{As Figure~\ref{fig:massres15b} but with $\beta=1.7$.}
\label{fig:massres17b}
\end{figure}
\begin{figure}
\includegraphics[width=\columnwidth]{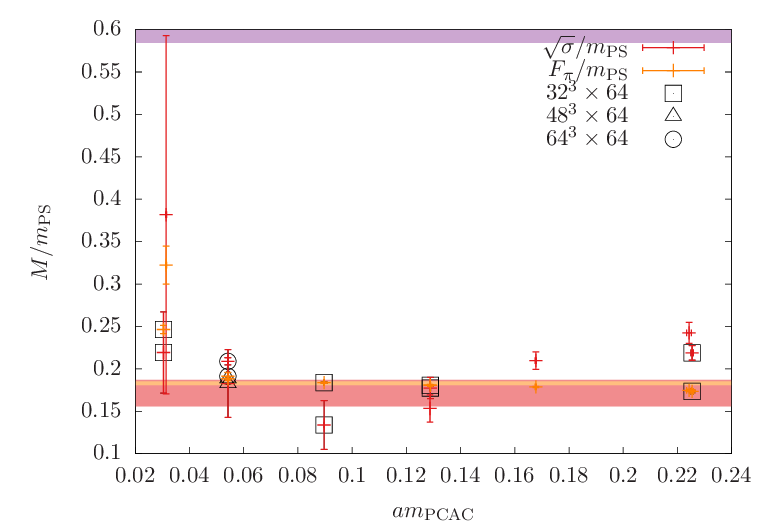}\\
\includegraphics[width=\columnwidth]{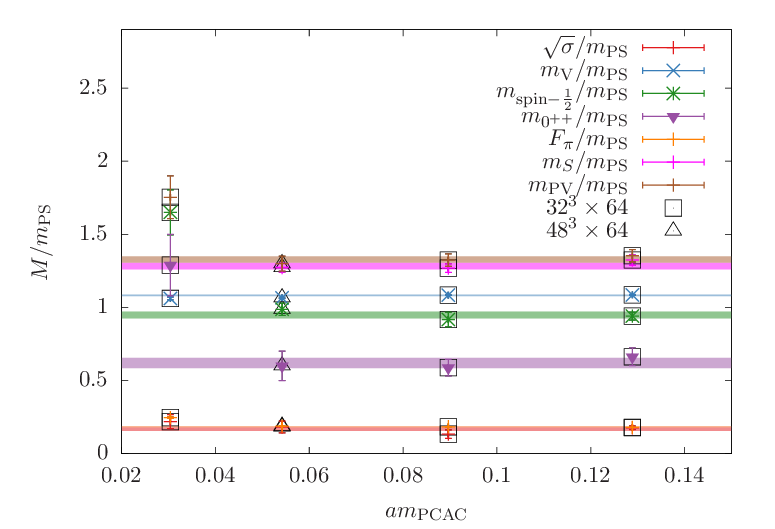}
\caption{Further details concerning the plot in Fig.~\ref{fig:massres15b}. 
These Figures now include the data at $\kappa=0.1350$ too. The upper 
figure shows the string tension and the pseudoscalar decay constant. In the
lower Figure the complete set of ratios is shown in the lower mass
region.}
\label{fig:massres15d}
\end{figure}

The results for the masses of the different states are shown in
Figs.~\ref{fig:massres15a}, \ref{fig:massres15b}, \ref{fig:massres17a} and
\ref{fig:massres17b}. First of all, it can clearly be observed that the mass
hierarchy is different from the one in the chiral symmetry breaking
scenario. Instead of the would-be Goldstone particle ($\mps$) the scalar
($0^{++}$) glueball is the lightest state in the theory. Furthermore, the
ratios of different quantities divided by $\mps$ are not divergent in the
zero fermion mass limit, as it would be the case for Goldstone particles.
Instead, they approach approximately constant values. These observations are
consistent with an IR conformal behaviour of the theory.

At the smallest fermion masses, in particular for $\kappa=0.135$ at
$\beta=1.5$, the results deviate significantly from the expected constant
mass ratios. The vector meson mass ratio is relatively stable, but for the
glueball even an inversion of the mass hierarchy with $\mps<\mgb$ is
observed. The ratios of string tension and pseudoscalar decay constant over
$\mps$ raise in this region, see Fig.~\ref{fig:massres15d}. A similar
observation has been made for the string tension in \cite{DelDebbio:2010hx},
where it has been traced back to a finite size effect at small $\mps L$. We
therefore conclude that the inverted mass hierarchy at these small fermion
masses is not a physical feature, but rather a finite size effect. A more
detailed investigation of this effect would require simulations on very
large lattices, which is beyond the scope of our current investigations. In
the estimates of the mass ratios we have therefore excluded the run at
$\beta=1.5$, $\kappa=0.135$ and, for the same reason, the run at
$\beta=1.7$, $\kappa=0.130$. Note that the clear distinction of the chiral 
symmetry breaking and the conformal scenario is quite challenging due to the 
limited accessible parameter space. A chiral symmetry 
breaking would be indicated by a $1/\sqrt{\mpcac}$ divergence of the mass ratios 
at small fermion masses. Out data do not favour this scenario, in particular by the 
precise values for $\mv$, but it is not excluded that it becomes dominant 
at even smaller fermion masses. The heavy meson mass ratios $\ms/\mps$ and $\mpv/\mps$ might
in addition have a subleading linear behaviour that we are not able to 
determine reliably.

\begin{table*}[ht]
\begin{center}
\begin{tabular}{ |c|c | c|c|}
\hline 
State & $\beta=1.5$ & $\beta=1.7$ & Ref.~\cite{DelDebbio:2015byq}\\
\hline
$\mv$ & 1.0825(58) & 1.051(12) & 1.044(43)\\
$\ms$ & 1.285(24)  & 1.190(14) & 1.222(52) \\
$\mpv$ & 1.329(21)  & 1.232(13) & 1.26(35) \\
$\mgb$ & 0.620(35)  & 0.398(48) & 0.458(15)\\
$F_\pi$ & 0.1831(23) & 0.15156(72) & 0.178(5)\\
$\sqrt{\sigma}$ & 0.171(16) & 0.1327(44) & 0.0959(14) -- 0.1319(10) \\
$\msh$ &  0.948(24) & 0.86394(52) & --\\
$\mpcac$ range & 0.1808(22)  -- 0.2490(12)  & 0.2457(12) -- 0.26776(42) & 0.1872(84) -- 0.2323(35) \\
$a\mps$ range & 0.29986(46) -- 0.58848(98) & 0.5360(25) -- 0.57247(16) & 0.6401(11) -- 1.183(1)\\
\hline
\end{tabular}
\end{center}
\caption{The masses of the different states in units of the pseudoscalar
mass $\mps$. The estimation is based on the approximate independence of
$\mpcac$. At $\beta=1.5$ the results from simulations on $32^3\times 64$ and
$48^3\times 64$ lattices with a $\kappa$ range between 0.1325 and 0.1344 are
taken into account. The results for $\beta=1.7$ are obtained from the
average of the $\kappa=0.1285$ and $\kappa=0.1290$ simulations on a
$32^3\times 64$ lattice. For comparison the results of
\cite{DelDebbio:2015byq} are shown, where for each state we have taken the
result at the smallest value of $m_0$. $F_\pi$ is an estimate from a plot in
\cite{DelDebbio:2015byq} and for the string tension we have shown the values
for two different $m_0$ since there are considerable deviations. In the
last line also the range of the reference scale $\mps$ in lattice units is
provided. Note that $F_\pi$ corresponds to the unrenormalised bare value.}
\label{tab:massrat}
\end{table*}

In spite of the limitations on the mass ranges, we are able to provide
estimates for the universal ratios between the different observables and the
pseudoscalar meson mass $\mps$. These are based on at least two values of
the mass for each $\beta$. The results are shown in Table~\ref{tab:massrat},
where also the results of \cite{DelDebbio:2015byq} are presented for
comparison. The general ordering of the masses at both $\beta$ values is
preserved, but all the ratios of masses to $\mps$ decrease from $\beta=1.5$
to $\beta=1.7$. For the meson masses the changes are below $8\%$, and for
the \shalf state they are slightly larger. The glueball mass, however, gets
corrections of the order of $50\%$. Consequently, the gap between the
glueball and the meson masses is significantly increased at the larger
$\beta$ value. This considerable difference cannot be explained by the
slightly different range of $\mpcac$ in units of $\mps$ for $\beta=1.5$ and
$\beta=1.7$. In particular, the $a\mps$ range is consistent for the two
$\beta$ values. Therefore we conclude that towards the continuum limit the
difference between the scalar glueball and the rest of the spectrum is in
fact increased.

We would like to note that the data in \cite{DelDebbio:2015byq} are all
approximately between our results at $\beta=1.5$ and $\beta=1.7$. Hence it
seems that our results at the coarse lattice spacing have larger lattice
artefacts, while the finer lattice spacing is closer to the continuum limit
than in \cite{DelDebbio:2015byq}. The string tension deviates from this
observation, but the values provided in the literature show a considerable
variance and hence seem to be subject to significant systematic
uncertainties. The uncertainties in our measurements are indicated by the
broad plateau estimation of the mass ratio, see Fig.~\ref{fig:massres15d}.
Further details of our measurement can be found in Section~\ref{sec:gbstr}.

Our arguments concerning the continuum limit in this section are based
simply on the asymptotic freedom of the gauge theory, which implies a
decreasing lattice spacing approaching $\beta=\infty$. If we assume that the
theory is already close to the conformal fixed point, the differences
between the results in Table~\ref{tab:massrat} are an indication for scaling
corrections. These effects seem to be relevant and to produce a significant
correction for the mass ratio of the scalar glueball and the pseudoscalar
meson mass.

\subsection{Scalar glueball and string tension}
\label{sec:gbstr}

The $0^{++}$ glueball appears to be the lightest scalar particle in aQCD2.
Despite a possible mixing with the scalar singlet meson operators, it seems
to have a reasonable overlap with the ground state in the scalar sector as
will be detailed in Section~\ref{sec:singmeson}. Hence it provides the
signal for a possible Higgs-like bound state. In some cases we have
also obtained an estimate for the $0^{-+}$ glueball mass (see
Table~\ref{tab:singmass}). It appears to be lighter than the pseudoscalar
singlet meson, but the systematic uncertainties are quite large.

We determine the $0^{++}$ glueball mass using as interpolating operator the
fundamental plaquette built from four links and the $0^{-+}$ is given by 
the product of eight links with suitable shape. In order to reduce the
contamination from excited states and thus determining the effective mass
already at small time-slice separations we used the variational method based
on APE smeared operators. In total, between $L = 16$ and $L = 20$ smearing
levels were used in the variational method, each separated by 4 or 5 steps;
the smearing parameter was fixed to $\epsilon_{APE}=0.5$.

\begin{figure}
\centering
\includegraphics[width=\columnwidth]{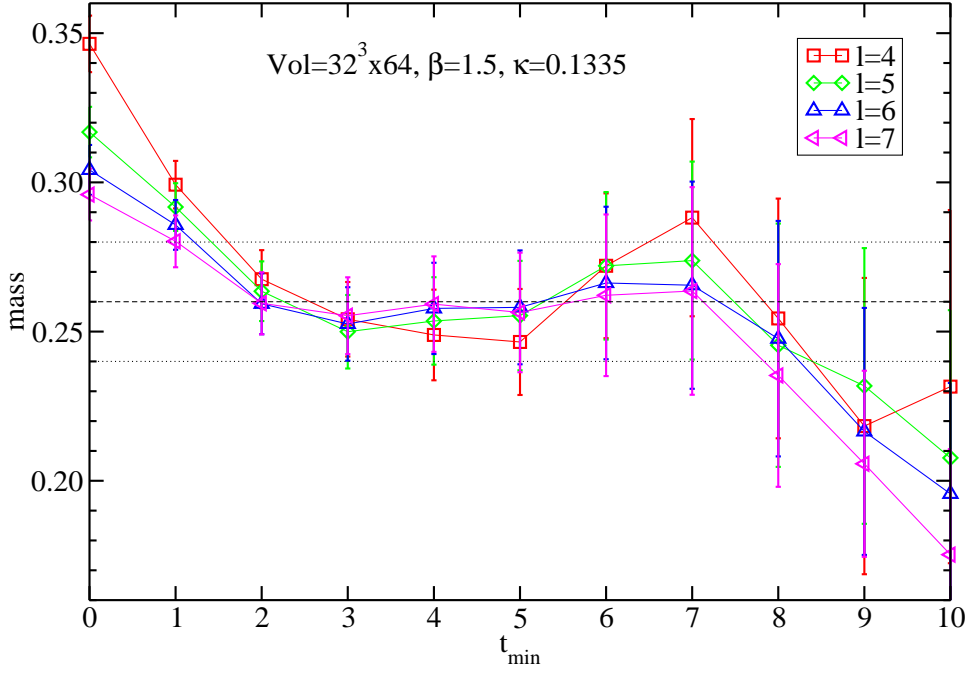}
\caption{Mass of the $0^{++}$ glueball obtained from fitting time slice
correlation functions in a range of time separations
$[t_{\text{min}},t_{\text{min}}+l]$.}
\label{fig:glub}
\end{figure}

Fig.~\ref{fig:glub} shows an example of the fitted mass value for different
ranges $[t_{\text{min}},t_{\text{min}}+l]$. A clear plateau appears already at
$t_{min}=2$. Using this approach we can determine the mass value with a
relative error starting from $10\%$ for some ensembles.

\begin{figure}[htb]
\begin{center} 
\includegraphics[width=\columnwidth]{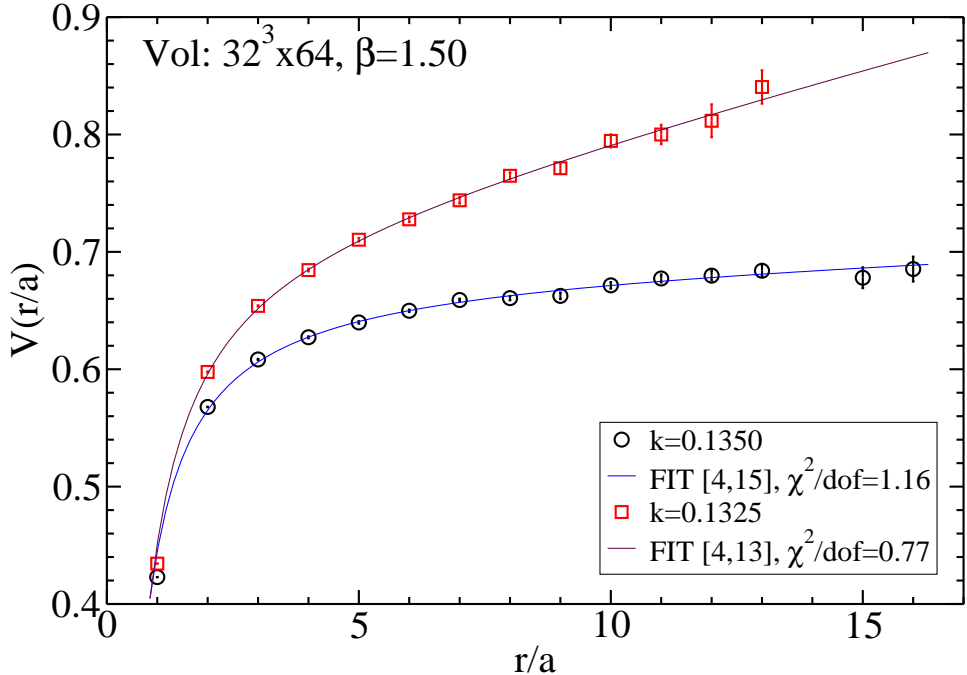}
\caption{Static quark-antiquark potential for two values of $\kappa$. At
large values of $r/a$ a few points are missing because of very large error
bars.}
\label{fig:sigma}
\end{center}
\end{figure}

For a theory in the confined phase, the potential between a static
quark-antiquark pair in the fundamental representation grows linearly at
large separations. The coefficient of the linear rise is the string tension
$\sigma$. In a theory with adjoint matter the chromoelectric field is not
screened and $\sigma$ is a well defined quantity. In an IR conformal theory the
string tension will vanish in the limit of massless fermions.

We determine the string tension from the expectation value of the Wilson loop
$\langle W(r,t) \rangle$. To this purpose we first define the 
generalised potential:
\begin{equation}
V(r,t)=\ln{\frac{\langle W(r,t) \rangle}{ \langle W(r,t+1) \rangle }}\, .
\end{equation}
The method consists of two steps: in the first one we determine the static
quark-antiquark potential fitting $V(r,t)$ for every $r$, in the interval
$[t_{\text{min}},t_{\text{max}}=L_T/2-1]$, to the function~\cite{Bali:1994de}
\begin{equation}
V(r,t)=V(r)+c_1 e^{-c_2 t}\, ;
\end{equation}
in the second step we fit the potential $V(r)$ to the form of the Cornell
potential, in the interval $[r_{\text{min}},r_{\text{max}}=L_S/2-1]$, and we
determine the value of the string tension. We have verified that, compared
to supersymmetric Yang-Mills theory, the value of $t_{\text{min}}$, in the
first fit, has to be increased from the value 2 to the value 3 and the value
of $r_{\text{min}}$, in the second fit, has to be increased from 2 to 4. As
a consequence, the potential $V(r)$ is characterised by large error bars, in
particular at large $r/a$ as can be seen in Fig.~\ref{fig:sigma}, and the
final string tension has a relative error $\sim 10$ times larger than the
case of supersymmetric Yang-Mills theory.

\subsection{Singlet meson states and a second signal for the scalar channel}
\label{sec:singmeson}

In our work we investigated the singlet meson sector of adjoint QCD for the
first time. The measurement of these states is more challenging than for the
rest of the spectrum, because their correlation functions contain
disconnected fermion contributions. For the calculation of these
contributions we have used the same methods that we have already applied in
our studies of supersymmetric Yang-Mills theory. It turns out that there are
significant systematic uncertainties in the measurement of these states and
therefore the reported errors are most likely underestimated. The results
for the masses are shown in Table~\ref{tab:singmass}.

The primary aim of these investigation is to obtain a second signal for the
scalar channel. The $\afn$ has the same quantum numbers as the $0^{++}$
glueball, and hence both operators have overlap with the ground state in the
scalar channel. The overlap with the ground state might, however, be small
such that a significant contribution from excited states is present. In the
case of supersymmetric Yang-Mills theory and one-flavour adjoint QCD it
turned out that there is a reasonable agreement between the two signals in
the scalar channel.

\begin{table*}
\begin{center}
\begin{small}
\begin{tabular}{|c|c|c|c|c|c|c|c|c|c|}
\hline
$L_S$ &$L_T$ &$\beta$ &$\kappa$ &$a m_{\afn}$&$a m_{\aetap}$ & $a\mps$&$a\ms$ &$a\mgb$ &$a\mgbmp$ \\
\hline
24 & 64 & 1.5 & 0.1325 &0.511(53)& 0.634(14) & 0.58710(27) & 0.767(12)&0.350(25)  & 0.62(9)\\
\hline
32 & 64 & 1.5 & 0.1335 &0.295(70)& 0.474(49)& 0.44212(28) &0.561(12)&0.260(25) & 0.44(5) \\
\hline
48 & 64 & 1.5 & 0.1344 & 0.320(69)& 0.342(36) &0.29986(46)&0.3816(87)&0.180(30) & 0.32(4)\\
\hline
32 & 64 & 1.7 & 0.1285 &0.515(52)& 0.574(28) &0.57247(16)&0.6902(62)&0.230(25) & 0.43(1) \\
\hline
32 & 64 & 1.7 & 0.1290 &0.419(66)&0.504(31) &0.5360(25)&0.6312(23)&0.210(30) & 0.36(2) \\
\hline
\end{tabular}
\end{small}
\end{center}
\caption{This Table contains the masses for the singlet mesons $\afn$ and
$\aetap$ in lattice units and, for comparison, also for some triplet mesons
and the glueballs.}
\label{tab:singmass}
\end{table*}

\begin{figure}
\includegraphics[width=\columnwidth]{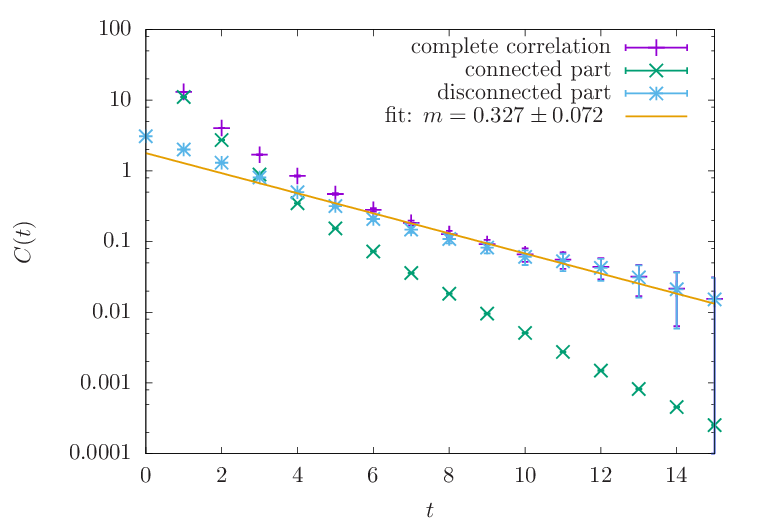}\\
\includegraphics[width=\columnwidth]{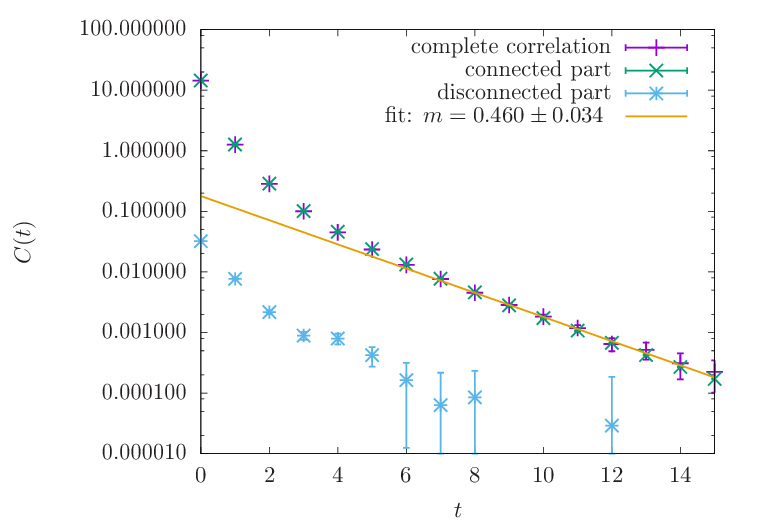}
\caption{In this Figure the disconnected and connected contributions to the
$\afn$ (above) and $\aetap$ (below) correlators are
shown. The sum of both parts gives the final correlation function. The mass
is obtained from the exponential decay of this function at large distances.}
\label{fig:disc}
\end{figure}

In aQCD2, the measurement of the connected and disconnected contributions of
the correlators leads to quite different results in the scalar and the
pseudoscalar case. While for the $\aetap$ meson the disconnected
contribution is almost negligible, see Fig.~\ref{fig:disc} (rhs), it is the
dominant contribution for the $\afn$, as shown in Fig.~\ref{fig:disc} (lhs).
The large disconnected contribution is the reason for the large difference
between $\ms$ and $m_{\afn}$. The scalar singlet meson mass ($m_{\afn}$) is
consequently much smaller than the mass in the triplet channel ($\ms$).
Nevertheless, there is no degeneracy with the light scalar glueball. The
mass of the scalar singlet meson is of the order of the pseudoscalar meson
mass $\mps$ or even lighter. It is interesting that in several
investigations of near IR conformal theories a similar approximate degeneracy
between the scalar singlet and the pseudoscalar meson has been
observed~\cite{Brower:2015owo,Fodor:2015vwa}.

Our results are an indication that the ground state in the scalar channel is
dominated by the glueball state. An analysis of the mixing between the
glueball and the $\afn$ meson will provide further information about the
overlap of the different operators with the lightest state.

Taking into account the systematic error in the evaluation of the singlet
mesons, the $\aetap$ meson is almost degenerate with the pseudoscalar meson.
The difference between these states, which in QCD is related to the axial
anomaly, is negligible within the current precision.

\subsection{Spin one-half states and possible fractionally charged particles}

A specific feature of theories with fermions in the adjoint representation
of the gauge group is the presence of mixed fermion-gluon states, which do
not occur in QCD. The most interesting one is a fermionic \shalf particle
represented by the operator
\begin{equation}
O_{\text{spin-1/2}} 
= \sum_{\mu,\nu} \sigma_{\mu\nu} \Tr\left[F^{\mu\nu} \lambda \right].
\end{equation}
This particle is of particular importance in supersymmetric Yang-Mills
theory, where this gluino-glue particle is the fermionic member of the
scalar supermultiplet of bound states. Unbroken supersymmetry implies a
degenerate mass for all the states of the supermultiplet, and hence it has
the same mass as the lightest scalar and pseudoscalar particle in this
model.

In aQCD2 the \shalf state is relevant for phenomenological considerations,
since it leads to fractionally charged particles, when a naive hypercharge
assignment is assumed. Even though the mass of these particles is unknown,
they have been considered to disfavour the phenomenological relevance of the
theory. This was essentially one of the motivations to consider \so{4} gauge
theory as an alternative \cite{Hietanen:2013gva}. On the other hand, in
\cite{Kouvaris:2007iq} the existence of such particles has been considered
as an alternative dark matter scenario.

Our results for the mass of the \shalf state are contained in Table
\ref{tab:massrat}. They show that the mass of the \shalf state is well
separated from the lightest scalar particle. On the other hand, it is
slightly lighter than the pseudoscalar meson, which means that it could be
one of the first experimentally observable ``new physics'' states if this
theory is realised in nature.

\section{The mass anomalous dimension}
\label{sec:gam}

Besides the ratios of different observables, also the mass anomalous
dimension $\gamma^\ast$ is an important universal property of an IR conformal
gauge theory. Given the ratios and the value of $\gamma^\ast$, the main
properties of the theory are determined. The mass anomalous dimension is of
particular importance for phenomenological considerations, since in the
\textit{Walking Technicolor} scenario a large value of $\gamma^\ast$ is
required. We apply two different methods to determine of $\gamma^\ast$. They
are based on the properties of particle spectrum and of the mode number.

\subsection{Scaling of the particle spectrum}

\begin{table}
\begin{center}
\begin{tabular}{ |c|c | c| }
\hline 
Observable & $\beta$ & $\gamma^\ast$\\
\hline 
$\mps$ &1.5 &0.2958(45)\\
$\mv$  &1.5 &  0.295(26)\\
\hline 
$\mps$ &1.7 & 0.289(17)\\
$\mv$  &1.7 &  0.263(28)\\
\hline
\end{tabular}
\end{center}
\caption{The values of the mass anomalous dimension determined from the fit
of $\mps$ and $\mv$. These results are based on a linear fit in a
double logarithmic representation. At $\beta=1.5$ only the result on the
$32^3 \times 64$ and the $48^3 \times 64$ lattices without $\kappa=0.1350$
are considered. At $\beta=1.7$ the values on the $32^3 \times 64$ lattice
without $\kappa=0.1300$ are taken into account. A fit of the other 
states are excluded due to the large fit error.}
\label{tab:gammass}
\end{table}
\begin{figure}
\centering
\includegraphics[width=\columnwidth]{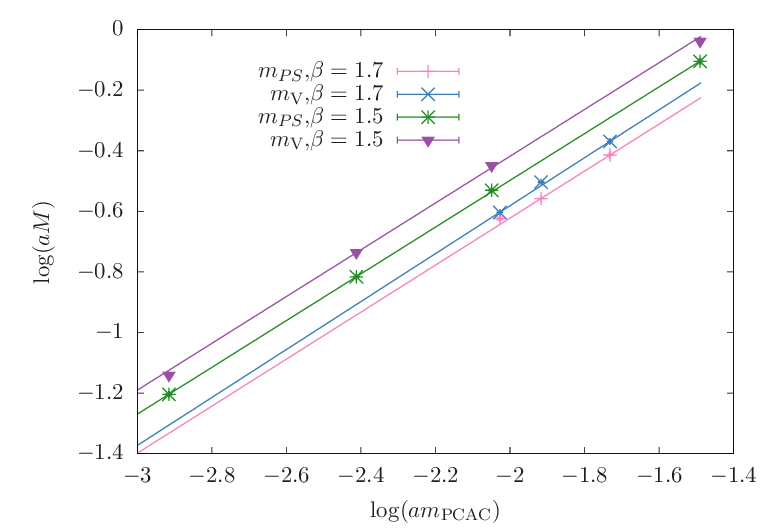}
\caption{This Figure illustrates the fit for the mass anomalous dimension
$\gamma^\ast$ from the masses $\mv$ and $\mps$. The considered subset of the data
and the fit results are given in Table~\ref{tab:gammass}. The figure above
corresponds to $\beta=1.5$, while the data below show the
results at $\beta=1.7$.}
\label{fig:gammass}
\end{figure}
\begin{figure}
\centering
\includegraphics[width=\columnwidth]{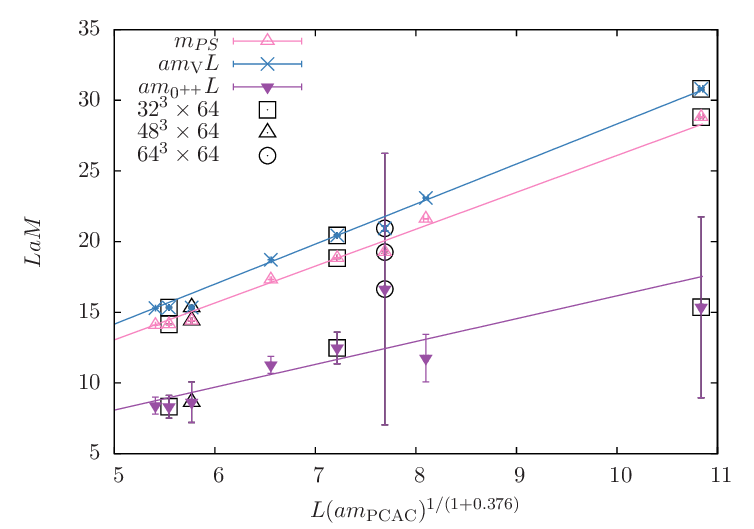}\\
\includegraphics[width=\columnwidth]{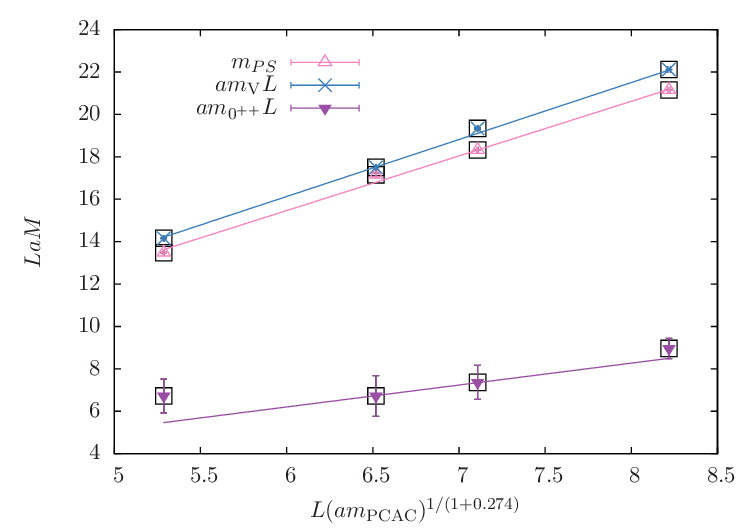}

\caption{This Figure shows a check for the consistency of the scaling of the
particle spectrum, with the mass anomalous dimension determined from the fit
of the mode number, see Table 4 for the values used. The figure above is
for $\beta=1.5$ and the one below for $\beta=1.7$. The approximate
volume scaling has been taken into account, where $L=N_s$ is the box size in
lattice units. The linear fit lines show the predicted approximate finite
size scaling relation.}

\label{fig:scalinggam}
\end{figure}

The fact that the masses of all states should scale according to the
universal formula
\begin{equation}
M \propto (\mpcac)^{1/(1+\gamma^\ast)}\, ,
\end{equation}
can be used to determine the mass anomalous dimension directly from the
particle spectrum. The simplest way to determine the exponent is a linear
fit in a double logarithmic representation. 
It turns out that for most of the masses the fit errors are large, and
the fit results spread over a range of $\gamma^\ast$ between 0.13 and 0.57.
For a first estimate it is thus reasonable to consider only the
most precise fits. Restricting the analysis to those states
that lead to a fit error smaller than 10\,\%, reduces the fit to
the pseudoscalar and vector meson mass.
The results of this fit are shown in Table~\ref{tab:gammass}
and Fig.~\ref{fig:gammass}. 
From these data one obtains a mass anomalous
dimension around $\gamma^\ast=0.3$, and there is a tendency towards a
smaller value at the larger $\beta$.

The large errors for several observables means that these fits represent
only a rough estimate of the mass anomalous dimension. A precise
determination of the mass anomalous dimension requires more control over the
parameter range. This can be achieved by fits of the mode number, where an
ultraviolet and infrared cutoff is introduced by the fit boundaries. The
consistency with the scaling of the spectrum can then be confirmed
subsequently. In the scaling formula also the approximate influence of the
finite volume can be taken into account by expressing the states in units of
the box size $L$. This scaling with the value of $\gamma^\ast$ obtained from
the mode number is shown in Fig.~\ref{fig:scalinggam} (see
Section~\ref{sec:moden}). As can be seen, within the current precision the
data of the particle spectrum are consistent with the scaling obtained from
the mode number. Note that in both cases the value of $\gamma^\ast=0.274$ is
preferred in comparison to $\gamma^\ast=0.376$ by the smaller chi-square in
the linear fit of $\mps$.

\subsection{Mode number}
\label{sec:moden}

\begin{table}[ht]
\begin{center}
\begin{tabular}{ |c|c | c| c| c| }
\hline 
$N_s\times N_t$ & $\beta$ & $\kappa$ & fit range & $\gamma^\ast$\\
\hline 
$24\times 64$ &1.5 & 0.1325 & 0.21-0.40 & 0.39(3) \\
$32\times 64$ &1.5 & 0.1335 & 0.21-0.40 & 0.38(1) \\
$48\times 64$ &1.5 & 0.1344 & 0.15-0.38 & 0.380(10) \\
$32\times 64$ &1.5 & 0.1350 & 0.11-0.37 & 0.375(4) \\
average           &1.5&&&0.376(3)\\
\hline 
$32\times 64$ &1.7 & 0.1285 & 0.38-0.57 & 0.270(15) \\
$32\times 64$ &1.7 & 0.1290 & 0.36-0.59 & 0.260(20) \\
$32\times 64$ &1.7 & 0.1300 & 0.28-0.50 & 0.285(15)\\
average            &1.7&&&0.274(10)\\
\hline
Ref.~\cite{Patella:2012da} & & & & 0.371(20) \\
\hline
Ref.~\cite{Perez:2015yna} & & & &  0.269(2)(5)\\
\hline
Ref.~\cite{Rantaharju:2015yva} & & & & 0.20(3)\\
\hline
Ref.~\cite{DeGrand:2011qd} & & & & 0.31(6)\\
\hline
Ref.~\cite{DelDebbio:2010hx} & & & & 0.22(6)\\
\hline 
Ref.~\cite{Giedt:2015alr} & & & & 0.50(26)\\
\hline
\end{tabular}
\end{center}
\caption{The mass anomalous dimension obtained from fits of the mode number.
For comparison we provide also some reference values from the literature
obtained with several different methods: Ref.~\cite{Patella:2012da} and
\cite{Perez:2015yna} are based on the mode number analysis. In
\cite{Perez:2015yna} this is done in a volume reduced large $N_c$ gauge
theory. Ref.~\cite{Rantaharju:2015yva} and \cite{DeGrand:2011qd} apply a
Schr\"odinger functional analysis. Ref.~\cite{DelDebbio:2010hx} and
\cite{Giedt:2015alr} use finite size scaling for the determination of the
mass anomalous dimension.}
\label{tab:modenumber}
\end{table} 
\begin{figure}
\centering
\includegraphics[width=\columnwidth]{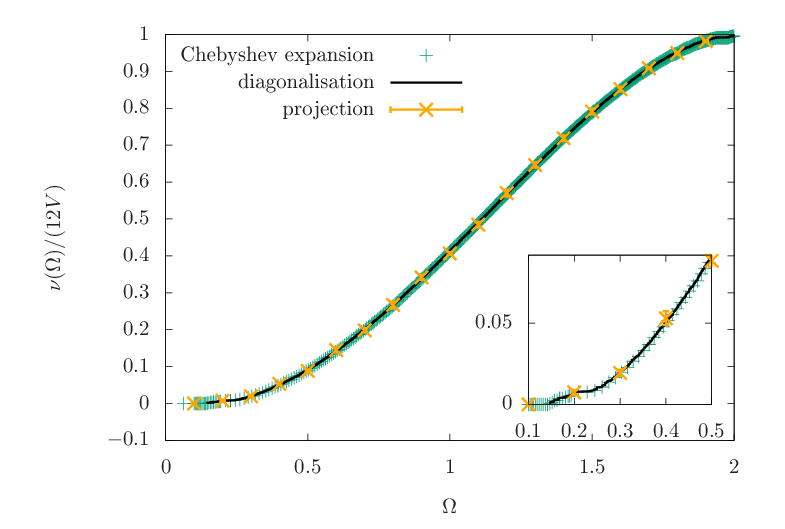}
\caption{This Figure shows a test of the projection method and the Chebyshev
expansion method for the determination of the mode number. The two methods
are compared to the numerical diagonalisation of the complete matrix on a
$4^4$ lattice for one configuration. In case of the projection method we
also included the stochastic error from the five noise vectors. The
Chebyshev expansion is based on a polynomial of order $N_p=1000$ and
$N_s=10$ estimators.}
\label{fig:modenumberchecks1}
\end{figure}
\begin{figure}
\centering
\includegraphics[width=\columnwidth]{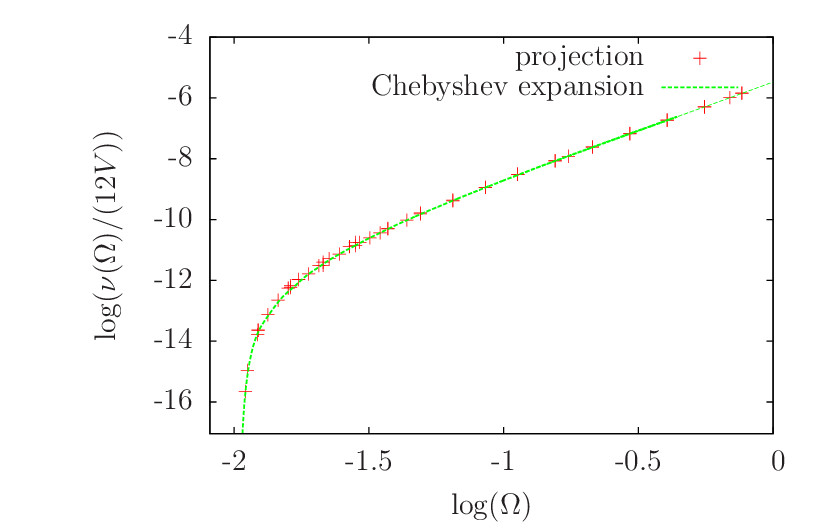}
\caption{The comparison of the projection method and the Chebyshev expansion
method for the determination of the mode number for one configuration on a
$32^3\times 64$ lattice ($\beta=1.7$, $\kappa=0.1290$) with $N_p=2000$,
$N_s=10$ for the Chebyshev expansion and only one estimator for the
projection method. The projection method is based on a polynomial
approximation of order 32.} \label{fig:modenumberchecks2}
\end{figure}

The mode number, which is the integrated eigenvalue density of the Dirac
operator, allows for a more precise estimate of the mass anomalous dimension
\cite{Patella:2012da,Cheng:2013bca,Fodor:2014zca}. On the lattice the most
practicable definition is obtained from the spectral density of the Dirac
operator. Let
\begin{equation}
\rho(\omega)=\frac{1}{V}\sum_k \langle \delta(\omega-\lambda_k) \rangle
\end{equation}
be the spectral density of the massless Dirac operator. The mode number
$\nu(\Omega)$, defined to be the number of eigenvalues of the
positive-definite operator $D_w^\dag D_w$ below some limit $\Omega^2$, is
given by
\begin{equation}
\nu(\Omega)=2\int_0^{\Lambda} \rho(\omega)d\omega\, ,
\end{equation}
where the cutoff for the integral is $\Lambda=\sqrt{\Omega^2-m_R^2}$, and
$m_R$ corresponds to the renormalised quark mass, i.\,e.\ is proportional to
$\mpcac$. Finally, the mass anomalous dimension is obtained from a fit of
the mode number (see \cite{Patella:2012da}) according to
\begin{equation}
\label{eq:modenumber}
\nu(\Omega)=a_1+a_2(\Omega^2-a_3^2)^{2/(1+\gamma^\ast)}\; .
\end{equation}
The constant $a_1$ is expected to scale like $\mps^4$, and $a_3$ is
proportional to $\mpcac$, but for our considerations these constants are not
relevant.

The projection method used in the earlier investigations of the mode number
was first proposed in \cite{Giusti:2002sm}. It is based on a rational
approximation of the projection operator $\Pro$ in the region below a
certain threshold of the eigenvalues. The mode number is hence defined as
\begin{align}
\nu(\Omega) = \langle\Tr \Pro(\Omega) \rangle \, ,
\end{align}
where the trace is obtained by a stochastic estimate. The projection
operator is approximated by means of a polynomial approximation of the step
function $h(x)$ using 
\begin{equation}
\Pro(\Omega)\approx h(\mathbb{X})^4\; , \qquad\text{with}\quad
\mathbb{X}=1-\frac{2\Omega_\ast^2}{D_w^\dag D_w+\Omega_\ast^2} \, .
\end{equation}
The parameter $\Omega_\ast\approx \Omega$ is adjusted in such a way that the
error of the approximation is minimised (see \cite{Giusti:2002sm} for
further details).

More recently a different method, based on a Chebyshev expansion of $\rho$,
has been proposed in \cite{Fodor:2016hke}. We have mainly used a variant of
this method, but we also checked the consistency with the projection method.
For the Chebyshev expansion method the spectrum has to be rescaled to the
interval $[-1,1]$ according to
\begin{align}
M=\frac{2D_w^\dag  D_w - \lambda_{max}-\lambda_{min}}{\lambda_{max}-\lambda_{min}}\; ,
\end{align}
where $\lambda_{max}$ and $\lambda_{min}$ are the maximal and minimal
eigenvalues of the operator $D_w^\dag D_w$. The integral of the spectral
density $\rho_M$ of the rescaled operator multiplied by the Chebyshev
polynomial $T_n$ of order $n$
\begin{align}
c_n=\int_{-1}^{1} \rho_M(x) T_n(x) \, ,
\end{align}
is estimated stochastically with $N_S$ random $Z_4$ noise vectors $v_l$:
\begin{align}
c_n\approx \frac{1}{N_S} \sum_l^{N_s} \langle v_l| T_n(M) | v_l \rangle .
\end{align}
Based on the orthogonality relations for the $T_n$, the spectral density
$\rho_M$ is now approximated by
\begin{align}
\rho_M(x) \approx 
\frac{1}{\pi \sqrt{1-x^2}} \sum_{k=0}^{N_p} (2-\delta_{k0}) c_n T_n(x).
\end{align}
The eigenvalue density of $D_w^\dag D_w$ is obtained from a simple map of the 
the interval $[-1,1]$ back to the original eigenvalue region.

The integral in the definition of the mode number can be performed
analytically. In our measurements we considered polynomials of order $N_p$
between 2000 and 4000.
\begin{figure}
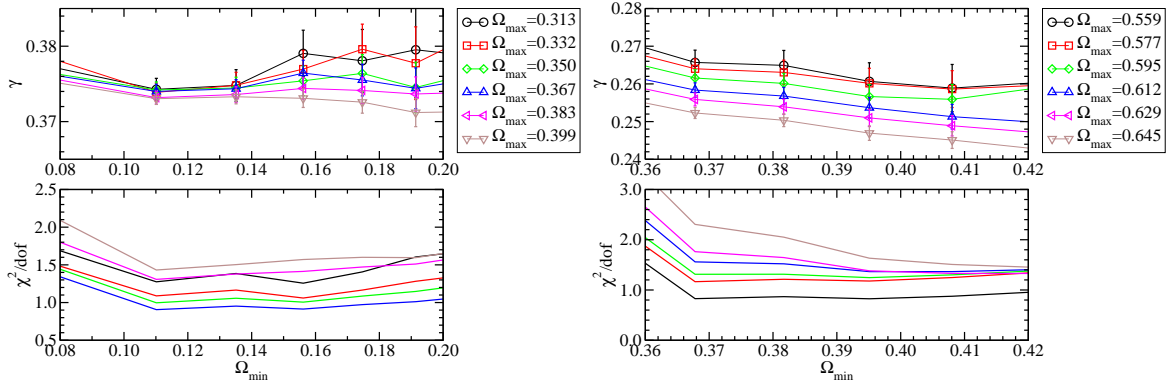

\includegraphics[width=\columnwidth]{plot_Run32c64_StoutB150C083k1350.eps}\\
\includegraphics[width=\columnwidth]{plot_Run32c64_StoutB170C083k1290.eps}
\caption{Results for $\gamma^\ast$ and $\chi^2/\mbox{dof}$ for different fit
ranges $[\Omega_{\text{min}},\Omega_{\text{max}}]$. 
Left: lattice $32^3 \times 64$, $\beta=1.5$, $\kappa=0.1350$. 
Right: lattice $32^3 \times 64$, $\beta=1.7$, $\kappa=0.1290$.}
\label{fig:modenfit}
\end{figure}
As a check we have compared the results of the two methods with the mode
number obtained from a complete numerical diagonalisation of $D_w^\dag D_w$
on small lattices. In addition we performed measurements with both methods
on a small number of configurations on $24^3\times 64$ lattices. The results
of these checks are shown in Figures~\ref{fig:modenumberchecks1} and
\ref{fig:modenumberchecks2}. Additional investigations and comparisons will
be done in the future for further understanding of the different methods. At
the moment, for the present measurements of the anomalous dimension based on
a limited range of $\Omega$ both methods are compatible.

The fitting procedure of the mode number $\nu(\Omega)$ deserves special
mention. Because the fitting data are strongly correlated, to determine
correctly the value of $\chi^2/\mbox{dof}$, we used the usual $\chi^2$
method, taking into account the correlation matrix. As discussed
in~\cite{Michael:1993yj}, to estimate correctly the value of
$\chi^2/\mbox{dof}$, the square of the number of fitted data has to be
smaller than the number of configurations used. Because the mode number is
measured on a number of configurations ranging in the interval $[100,1000]$,
the number of fitted points ranges in the interval $[10,30]$. For
comparison, the fitting parameters have been also determined by means of
uncorrelated fits, using in this case a number of fitted data of the order
of the number of configurations~\cite{Michael:1994sz}, giving compatible
results.

Another issue in the fitting procedure is related to the fact that
Eq.~(\ref{eq:modenumber}) can be used only in a certain intermediate range
of eigenvalues, that can be determined only by a systematic study of the
quality and the stability of the fit. As shown in Fig.~\ref{fig:modenfit},
we fit the data for different values of the range
$[\Omega_{\text{min}},\Omega_{\text{max}}]$, looking for values which
guarantee a plateau in $\chi^2/\mbox{dof} \sim 1$ and in the value of the
mass anomalous dimension $\gamma^\ast$.
\begin{figure}
\centering
\includegraphics[width=\columnwidth]{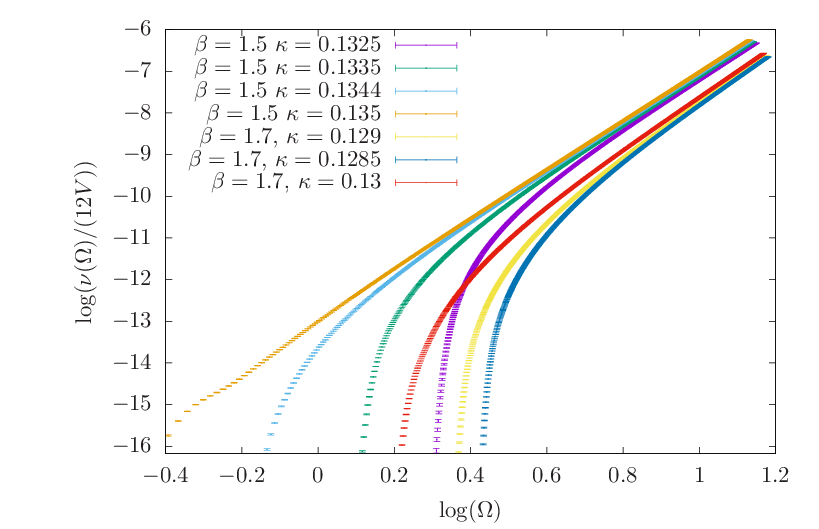}
\caption{The mode number data that have been used in the final fit to obtain
the results presented in Table~\ref{tab:modenumber}.}
\label{fig:modenumberdata}
\end{figure}
Our results for the mass anomalous dimension are presented in
Fig.~\ref{fig:modenumberdata}. The results of the fits are shown in
Table~\ref{tab:modenumber}. In this table we also considered the runs with
sizeable finite size effects, because we assume that these effects only
influence the far infrared region and not the part relevant for the fits.
The results for these runs are compatible with the other runs.

The values of $\gamma^\ast$ obtained via mode number are in reasonable
agreement with those from the mass spectrum, especially for the smaller
lattice spacing at $\beta=1.7$.

Our result at $\beta=1.5$ is consistent with \cite{Patella:2012da}, where
$\gamma^\ast=0.371(20)$ has been reported. The value at $\beta=1.7$,
however, appears to be significantly smaller. Thus there is still a
remaining $\beta$ dependence of the mass anomalous dimension, reducing its
value towards larger $\beta$ values. It is interesting to note that such
smaller values have also been reported in other works, for example in
\cite{Rantaharju:2015yva}.

Note that the estimates of $\gamma^\ast$ discussed in this work show a
remnant dependence on the gauge coupling, and are thus not precisely
identical with the universal value at the IR fixed point. The determination
of the value at the fixed-point would require to include scaling corrections
in the analysis, which is beyond the present possibilities.
\section{Conclusions}
\label{sec:concl}

In this work we have presented results for \su{2} gauge theory with two
flavours of Dirac fermions in the adjoint representation regarding the particle
spectrum and concerning general signals for conformality at two different
values for the inverse gauge coupling $\beta$ and several masses of the two
adjoint Dirac fermions. The structure of the low-lying spectrum of particle
masses shows clear indications for an IR conformal behaviour. This qualitative
observation was complemented with quantitative results for the universal
mass ratios and the mode number.

In earlier investigations the (triplet) mesonic spectrum and the glueballs have been considered.
Note in particular the detailed studies of finite volume effects in \cite{DelDebbio:2015byq} and of
the mode number in \cite{Patella:2012da}. As shown in Table~\ref{tab:modenumber}, these rather precise estimates 
for the mass anomalous dimension are not consistent with other investigations based on alternative methods.
Our results indicate that the consideration of the dependence on the gauge coupling $\beta$ and the corresponding 
scaling violation provides the missing link between these estimates.
At the smaller $\beta$ value we find, despite the differences in the lattice actions, 
results consistent with \cite{DelDebbio:2015byq} and \cite{Patella:2012da}.
At the larger $\beta$ value, however,
we find significantly lower results for the masses in units of the
pseudoscalar mass, especially for the glueball, and a smaller mass anomalous
dimension. Bearing in mind that different lattice actions have been used,
the unimproved one in \cite{Patella:2012da} and the clover improved fermion
action in \cite{Rantaharju:2015yva}, our results indicate that the mass
anomalous dimensions converge towards a universal value in the
continuum limit only, if possible scaling corrections are included. It seems
that towards that limit aQCD2 becomes even more conformal in the sense that
the gap between the scalar particle and the rest of the spectrum is
increased and the mass anomalous dimension gets smaller. In fact, our
results indicate that the differences between various numerical results for
the mass anomalous dimension are rather related to scaling corrections
than to the differences between the considered approaches.

Our work also provides a connection to the large $N_c$ results presented in
\cite{Perez:2015yna}. The mass anomalous dimension in conformal \su{N_c}
adjoint QCD is expected to depend only weakly on $N_c$. Therefore the
investigation in a large $N_c$ volume-reduced theory might be a valid
approximation. The mass anomalous dimension obtained in this approach is
consistent with our results at $\beta=1.7$.

In addition to these results, we have also been able to investigate particle
states that have not been considered before. The one with the most
interesting phenomenological consequences is the \shalf state. We have found
that it is considerably lighter than the mesons, and therefore 
it might eventually be interesting from a phenomenological point of view.

The general picture of the particle spectrum in aQCD2 appears to be ordered
starting with low mass pure gluonic states (glueballs), followed by heavier
mixed fermion-gluon objects, and finally the rather heavy triplet mesons.

We have also measured the singlet mesons in the particle spectrum of aQCD2
for the first time. The most interesting state is the scalar singlet meson.
Having the same quantum numbers as the lightest scalar particle in the theory,
it may give additional information about it. 
Due to the dominance of the disconnected contributions, the mass
of this particle is comparable or even below that of the pseudoscalar meson,
the lightest meson in the triplet channel. This is similar to the
observations reported in other studies \cite{Brower:2015owo,Fodor:2015vwa}
of (near) IR conformal theories. However, the ground state in this channel
seems to be dominated by the gluonic contributions, and hence 
the effective mass in the meson channel is still higher than the one
in the glueball channel. 
Therefore it is
not sufficient to measure only the mesonic contribution to get a complete
picture for the lightest scalar in this theory.

\section*{Acknowledgments}

We thank Agostino Patella, Biagio Lucini, Anna Hasenfratz, and Kieran
Holland for interesting discussions.

The authors gratefully acknowledge the Gauss Centre for Supercomputing (GCS)
for providing computing time for a GCS Large-Scale Project on the GCS share
of the supercomputer JUQUEEN at J\"ulich Supercomputing Centre (JSC) and on
the supercomputer SuperMUC at Leibniz Computing Centre (LRZ). GCS is the
alliance of the three national supercomputing centres HLRS (Universit\"at
Stuttgart), JSC (Forschungszentrum J\"ulich), and LRZ (Bayerische Akademie
der Wissenschaften), funded by the German Federal Ministry of Education and
Research (BMBF) and the German State Ministries for Research of
Baden-W\"urttemberg (MWK), Bayern (StMWFK) and Nordrhein-Westfalen (MIWF).
Further computing time has been provided by the computer cluster PALMA of the
University of M\"unster.


%
\begin{appendix}
\onecolumngrid
\newpage
\newpage
\section{Data}
\label{ap:data}
\begin{table*}[h!]
\begin{center}
\begin{small}
\begin{tabular}{|c|c|c|c|c|c|c|c|c|}
\hline
 $L_S$ &$L_T$ &$\beta$ &$\kappa$ &$a^2\sigma$ &$a\mpcac$ &$aF_{PS}$ &$am_{PS}$ &$N_\text{configs}$ \\
\hline
24 &64 &1.5 &0.13 &0.0477(48)&0.22429(88)&0.1573(12)&0.90080(67)&2130 \\
\hline
24 &64 &1.5 &0.1315 &0.0229(22)&0.16775(25)&0.12891(44)&0.72149(44)&9900 \\
\hline
24 &64 &1.5 &0.1325 &0.0081(17)&0.128730(46)&0.10634(27)&0.58710(27)&9800 \\
\hline
24 &64 &1.5 &0.135 &0.0037(40)&0.03136(15)&0.0514(29)&0.1593(20)&1720 \\
\hline
32 &64 &1.5 &0.13 &0.0388(30)&0.22539(47)&0.1561(13)&0.90030(91)&1480 \\
\hline
32 &64 &1.5 &0.1325 &0.0109(15)&0.128840(55)&0.10617(40)&0.58848(98)&3448 \\
\hline
32 &64 &1.5 &0.1335 &0.0035(15)&0.089619(74)&0.08125(28)&0.44212(28)&5627 \\
\hline
32 &64 &1.5 &0.135 &0.00140(60)&0.030414(45)&0.04204(68)&0.17063(65)&4432 \\
\hline
48 &64 &1.5 &0.1344 &0.0030(13)&0.054202(59)&0.05645(47)&0.29986(46)&1122 \\
\hline
64 &64 &1.5 &0.1344 &0.00395(12)&0.05417(11)&0.0576(10)&0.3009(15)&418 \\
\hline
32 &64 &1.7 &0.1275 &0.00708(41)&0.17697(22)&0.09717(29)&0.66093(22)&5069 \\
\hline
32 &64 &1.7 &0.1285 &0.00605(21)&0.147091(22)&0.08690(16)&0.57247(16)&11901 \\
\hline
32 &64 &1.7 &0.129 &0.00461(20)&0.131717(22)&0.08007(14)&0.5360(25)&11891 \\
\hline
32 &64 &1.7 &0.13 &0.00366(34)&0.100878(47)&0.06591(23)&0.42116(32)&3941 \\
\hline
\end{tabular}
\end{small}
\vspace*{0.8cm}

\begin{small}
\begin{tabular}{|c|c|c|c|c|c|c|c|c|}
\hline
 $L_S$ &$L_T$ &$\beta$ &$\kappa$ &$am_{V}$ &$am_{1/2}$ &$am_{0^{++}}$ &$am_{S}$ &$am_{PV}$ \\
\hline
24 &64 &1.5 &0.13 &0.9622(12)&0.933(14)&0.490(70)&1.179(97)&1.288(18)\\
\hline
24 &64 &1.5 &0.1315 &0.77990(45)&0.727(19)&0.470(25)&0.983(14)&1.029(21)\\
\hline
24 &64 &1.5 &0.1325 &0.63742(40)&0.551(23)&0.350(25)&0.767(12)&0.793(21)\\
\hline
24 &64 &1.5 &0.135 &0.1802(43)&0.292(13)&0.190(30)&0.400(22)&0.38(11)\\
\hline
32 &64 &1.5 &0.13 &0.9628(25)&0.917(18)&0.48(20)&1.223(41)&1.358(29)\\
\hline
32 &64 &1.5 &0.1325 &0.6387(11)&0.554(15)&0.390(35)&0.779(19)&0.796(24)\\
\hline
32 &64 &1.5 &0.1335 &0.47937(56)&0.406(22)&0.260(25)&0.561(12)&0.585(18)\\
\hline
32 &64 &1.5 &0.135 &0.18111(91)&0.282(25)&0.220(35)&--&0.299(24)\\
\hline
48 &64 &1.5 &0.1344 &0.31963(99)&0.296(12)&0.180(30)&0.3816(87)&0.390(15)\\
\hline
64 &64 &1.5 &0.1344 &0.3272(27)&0.248(12)&0.26(15)&0.3700(72)&0.399(37)\\
\hline
32 &64 &1.7 &0.1275 &0.69117(32)&0.564(24)&0.280(15)&0.7967(72)&0.832(13)\\
\hline
32 &64 &1.7 &0.1285 &0.60436(94)&0.4945(92)&0.230(25)&0.6902(62)&0.7115(56)\\
\hline
32 &64 &1.7 &0.129 &0.54693(26)&0.464(22)&0.210(30)&0.6312(23)&0.6514(38)\\
\hline
32 &64 &1.7 &0.13 &0.44244(62)&0.394(24)&0.210(25)&--&0.479(48)\\
\hline
\end{tabular}
\end{small}
\end{center}
\caption{These two tables contain the raw data obtained from the simulations
at the two different $\beta$ values. All values are provided in lattice
units.}
\label{tab:data}
\end{table*}
\end{appendix}
\end{document}